\documentclass{article}
\usepackage{float}

\PassOptionsToPackage{numbers}{natbib}
\usepackage[preprint]{neurips_2025}

\usepackage{natbib}
\usepackage{enumitem}
\usepackage{float}

\usepackage[utf8]{inputenc}        
\usepackage[T1]{fontenc}           
\usepackage{hyperref}              
\usepackage{url}                   
\usepackage{booktabs}              
\usepackage{amsmath, amssymb}      
\usepackage{nicefrac}              
\usepackage{microtype}             
\usepackage{xcolor}                
\usepackage{graphicx}              

\usepackage{enumitem}
\usepackage[utf8]{inputenc}
\usepackage[T1]{fontenc}
\usepackage{hyperref}
\usepackage{url}
\usepackage{booktabs}
\usepackage{amsmath, amssymb}
\usepackage{nicefrac}
\usepackage{textcomp}
\usepackage{microtype}
\usepackage{xcolor}
\usepackage{graphicx}
\usepackage{soul}

\newcommand{\model}{\textsc{AstroCo}}
\newcommand{\rmse}{\ensuremath{\mathrm{RMSE}}}

\title{\model: Self-Supervised Conformer-Style Transformers for Light-Curve Embeddings}

\author{
  \textbf{Antony Tan}$^{1,2}$, \textbf{Pavlos Protopapas}$^{1}$,
  \textbf{M. Cádiz-Leyton}$^{3,4}$, \\
  \textbf{Guillermo Cabrera-Vives}$^{3,4,5,6}$,
  \textbf{C. Donoso-Oliva}$^{4,6}$,
  \textbf{I. Becker}$^{5,6,7}$ \\[0.5em]
  \begin{tabular}{@{}c@{}}
  \small
  $^{1}$ John A. Paulson School of Engineering and Applied Sciences, Harvard University, USA \\
  \small
  $^{2}$ Department of Biostatistics, Harvard T.H. Chan School of Public Health, USA \\
  \small
  $^{3}$ Department of Computer Science, Universidad de Concepción, Chile \\
  \small
  $^{4}$ Center for Data and Artificial Intelligence, Universidad de Concepción, Chile \\
  \small
  $^{5}$ Millennium Institute of Astrophysics (MAS), Santiago, Chile \\
  \small
  $^{6}$ Millennium Nucleus on Young Exoplanets and their Moons (YEMS), Chile \\
  \small
  $^{7}$ Department of Computer Science, Pontificia Universidad Católica de Chile, Santiago, Chile
  \end{tabular}
}

\begin{document}
\maketitle
\begin{center}
\small\textit{Accepted at the NeurIPS 2025 Workshop on Machine Learning and the Physical Sciences (ML4PS). Camera-ready version in progress.}
\end{center}

\begin{abstract}

We present AstroCo, a Conformer-style encoder for irregular stellar light curves. By combining attention with depthwise convolutions and gating, AstroCo captures both global dependencies and local features. On MACHO \(R\)-band, \textsc{AstroCo} outperforms Astromer v1/v2, yielding 70\%/61\% lower error and a relative macro-$F_1$ gain of $\sim$7\%, while producing embeddings that transfer effectively to few-shot classification. These results highlight AstroCo’s potential as a strong and label-efficient foundation for time-domain astronomy.

\end{abstract}

\section{Introduction}
Self-supervised learning has recently advanced representation learning for astronomical time series. 
In particular, Donoso-Oliva and collaborators introduced \emph{Astromer} and its successor \emph{Astromer 2} \cite{AstromerAandA,Astromer2}, transformer-based models \cite{Vaswani2017} that learn embeddings of stellar light curves through masked reconstruction. 
These approaches have demonstrated that foundation-style models can provide compact, label-efficient representations for variable star classification and related downstream tasks.

Despite their success, existing transformer encoders often treat every time step equally, limiting their ability to capture short-lived local phenomena such as dips, flares, or bursts. 
Moreover, they lack explicit mechanisms to regulate how noisy or distant measurements influence the overall representation. 
This reduces both reconstruction fidelity and downstream generalization.

To address these challenges, we introduce \textbf{\model}, a conformer-style\cite{Conformer} encoder that combines global self-attention with local depthwise convolutions and gating. 
Our design improves the balance between long-range dependency modeling and local feature extraction, while introducing adaptive mechanisms for information flow across layers. 
We demonstrate that \model{} produces more accurate reconstructions and more label-efficient embeddings than previous Astromer baselines, and transfers effectively to few-shot settings in downstream classification tasks.

\section{Architecture and Training}

Each encoder block in \model{} combines three sublayers:  
(i) multi-head self-attention to capture long-range dependencies,  
(ii) a depthwise convolutional sublayer equipped with Gated Linear Units\cite{Dauphin2017} (GLUs, which use a learned gate to modulate local feature flow), and  
(iii) a gated feed-forward network that adaptively controls information at the global representation level.  
These components are tied together with residual connections and layer normalization.

For input, magnitudes and uncertainties are concatenated with time embeddings and fused through a learned projection, avoiding the scale mismatch that can arise in additive fusion (raw magnitudes and times are summed directly). To generate compact sequence embeddings, we mix features across all layers—including the input—via trainable softmax weights, followed by masked mean pooling.

On the MACHO $R$-band dataset, \textsc{AstroCo-S} (5.9M parameters; trained 11.6h on 4×A100 GPUs) and \textsc{AstroCo-L} (15.2M parameters; trained 1.2d on 4×H200 GPUs) both achieve lower reconstruction error and higher macro-F1 classification scores than Astromer v1/v2(5.4M parameters; trained 3d on 4×A5000 GPUs). See Appendix~\ref{app:implementation} for full hyperparameters and training details.
Notably, the smaller model surpasses Astromer despite using fewer resources, while the larger variant leverages its capacity to set the strongest overall benchmark. 

\textbf{Contributions.} Our design shows that (1) adding locality-aware convolutions and gating mechanisms greatly improves representation quality over purely attention-based encoders, (2) soft feature mixing across layers\cite{tenney2019bert} yields better embeddings than fixed pooling strategies, and (3) the approach is resource-efficient, with the small model outperforming prior work using less training time and hardware.

To realize these improvements, \model{} integrates attention, convolutions, and gating in a conformer-style encoder, which we outline next.

\section{Method}\label{sec:method}

\subsection{Inputs, embeddings, and masking}
\label{sec:inputs}
We model each light curve as an irregular sequence 
$\{(t_i, m_i, \sigma_i)\}_{i=1}^L$, where $t_i$ is the time, $m_i$ the magnitude, and $\sigma_i$ a measurement uncertainty.  
To encode inputs, photometric values $(m_i, \sigma_i)$ are projected to a $d/2$-dimensional vector, while times $t_i$ are mapped to $d/2$ dimensions using sinusoidal embeddings. These two representations are concatenated and fused into a $d$-dimensional embedding by a linear layer, followed by a GeLU activation and LayerNorm.

For self-supervised pretraining, we adopt a masked-reconstruction strategy. At each sequence, we select a target set of positions $\mathcal{M}$ (“probed positions”), covering 50\% of the time steps. Among these, $\sim$30\% are masked (the true values of $m_i$ are hidden), 10\% are replaced with random values, and 10\% are left unchanged but still included in the loss (BERT-style) \cite{BERT}. For masked or padded tokens, we replace the raw $(m,t)$ pair with a zero placeholder before projection, ensuring that the network cannot trivially recover missing values from input leakage.

\begin{figure}[H]
    \centering
    \includegraphics[width=0.9\linewidth]{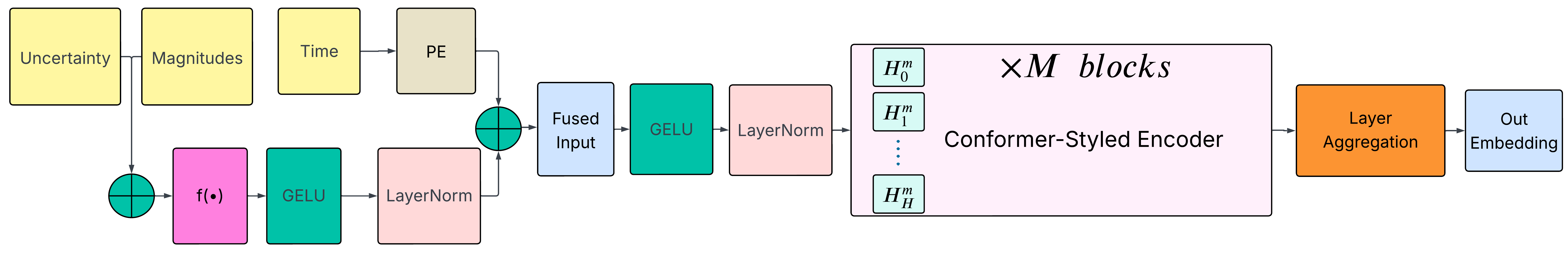}
    \caption{\textbf{\model{} architecture.} Magnitudes and uncertainties are projected and fused with sinusoidal time embeddings, then passed through a Conformer-style encoder. Hidden states from all layers are scalar-mixed to form the final sequence embedding (AstroCo-L: $M=12$, $H=4$, $m=64$, $D=256$).}
    
\end{figure}

\subsection{Encoder block architecture}

\begin{figure}[H]
    \centering
    \includegraphics[width=0.9\linewidth]{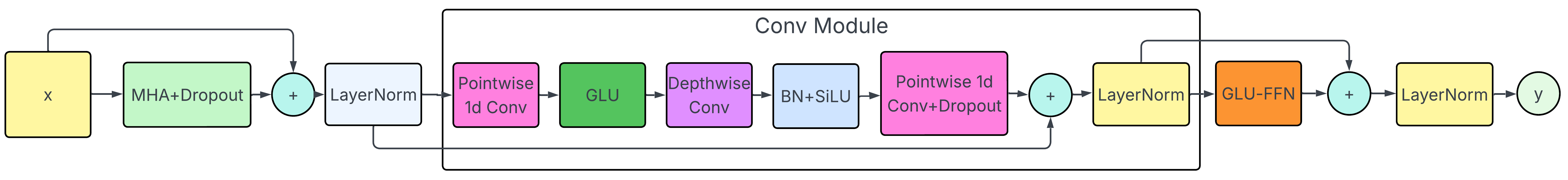}
    \caption{Encoder block consists of a MHSA layer, a depthwise convolutional sublayer with GLU gating, and a gated FFN, with residuals and LayerNorm throughout and BN + SiLU in the convolution.}
    
    \label{fig:arch}

\end{figure}

Each encoder block (Figure~\ref{fig:arch}) consists of three sublayers connected with residuals, dropout, and LayerNorm:  

\paragraph{(1) Multi-head self-attention. } 
Given $X \in \mathbb{R}^{B \times T \times D}$, we use standard MHSA then applied with dropout and residual followed by post-norm.

\paragraph{(2) Convolutional sublayer.}  
We first normalize the input ($\hat{X} = \mathrm{LN}(X)$), then apply a pointwise $1{\times}1$ projection $D \rightarrow 2D$ followed by a Gated Linear Unit (GLU), which controls the flow of local features. 
Next, a depthwise Conv1D with kernel size $K$ mixes information across time within each channel:
\[
Y[b,d,t] = \sum_{k=0}^{K-1} W^{\text{dw}}[d,k] \cdot X[b,d,\,t+k-\lfloor K/2 \rfloor],
\]
where $b$ indexes the batch, $d$ the channel, and $t$ the time step.  
This is followed by BatchNorm1d, a SiLU activation, and a final pointwise $1{\times}1$ projection back to $D$ \footnote{The $D \rightarrow 2D$ projection enables splitting into candidate values and gates, which the GLU combines as $\text{val} \odot \sigma(\text{gate})$. The subsequent depthwise convolution applies temporal filters per channel, capturing local patterns without cross-channel mixing.}. We then add the residual connection and apply LN.

\paragraph{(3) Gated feed-forward network (Gated FFN).}  
With gating expansion, $r=4$ \footnote{With expansion ratio $r$, the input is projected to two $rD$ vectors (via $W_{val}, W_{gate}$). After applying gating, a linear projection $W_{out}$ reduces back to dimension $D$.}, the FFN computes
\begin{align}
\text{val} &= \mathrm{GeLU}(W_{\text{val}} X), \quad
\text{gate} = \sigma(W_{\text{gate}} X), \\
Y &= X + \mathrm{Dropout}\!\big(W_{\text{out}}(\text{val}\odot\text{gate})\big), \quad
Y \leftarrow \mathrm{LayerNorm}(Y).
\end{align}
The gate adaptively modulates which global features are retained in the representation. And a residual addition and LN is applied before the encoder output.

\subsection{Layer-wise aggregation and pooled embeddings}
To obtain a single compact representation from the stacked blocks, we aggregate information across layers. Rather than relying only on the final layer, we learn a weighted combination of all intermediate representations, including the input. This allows the model to adaptively decide how much to emphasize shallow vs. deep features.

Let $x^{(0)}$ denote the input sequence and $x^{(\ell)}$ the output of block $\ell{=}1{:}M$. 
We learn scalar mixing weights $\{w_\ell\}$, normalized with a softmax $\alpha_\ell = \frac{\exp(w_\ell)}{\sum_{j=0}^{M}\exp(w_j)}.$
The final representation at position $i$ is the weighted sum over all layers (including the input): $\tilde{x}_i = \sum_{\ell=0}^{M} \alpha_\ell \, x^{(\ell)}_i.$

\subsection{Objective}\label{sec:objective}
Our training procedure consists of two distinct objectives, corresponding to the pretraining and downstream phases.  
During pretraining, the encoder is optimized to reconstruct photometric values by minimizing the root mean squared error (RMSE) between predicted and true magnitudes at the probed positions.  
For downstream classification, we freeze the pretrained encoder and train a linear head using cross-entropy loss (see Appendix~\ref{app:loss}). This setup ensures that the representations learned during pretraining can be leveraged for supervised tasks.

\section{Data and Setup}\label{sec:data}
\noindent \ul{Unlabeled pretraining:}

We pretrain on $\sim$1.5M single-band ($R$) light curves from the MACHO survey \cite{Alcock2003}. Following Astromer v2 settings \cite{AstromerAandA}, we first segment each series into length-200 windows and then normalize each window (zero-mean magnitude and time). Pretraining uses a BERT-style masking scheme \cite{BERT}: we select 50\% of positions as probed—of all positions, 30\% are replaced by masking scalar token, 10\% by a randomized value, and 10\% left unchanged—and compute the RMSE reconstruction loss on the probed points.

\noindent \ul{Labeled classification:}

For downstream evaluation, we use the MACHO LMC variable star catalog, which provides 20,894 labeled light curves across six broad classes \cite{Alcock2003}. 
To ensure label efficiency and comparability, we adopt the same labeled subset protocol as in prior Astromer work: for each class, 20, 100, or 500 instances are sampled without overlap across train, validation, and test splits. 
To reduce variance, we generate three independent folds for each setting, stratified by class. 
A linear head is then trained on top of the frozen encoder for each fold, and we report the mean macro-$F_1$ across folds.

\section{Results}\label{sec:results}
\subsection{Masked reconstruction}

\begin{table}[t]
\caption{\textbf{MACHO R-band: masked reconstruction results.} 
Lower is better for reconstruction \rmse{}; higher is better for $R^2$.}
\label{tab:mainresults}
\centering
\setlength{\tabcolsep}{6pt}      
\renewcommand{\arraystretch}{0.92} 
\footnotesize                      
\begin{tabular}{lcc}
\toprule
Model & Reconstruction \rmse{} $\downarrow$ & $R^2$ $\uparrow$ \\
\midrule
Astromer v1 (Donoso et al.\ (2023)) & 0.148 & -- \\
Astromer v2 (Donoso et al.\ (2025)) & 0.113 & 0.73 \\
\textsc{AstroCo-S}    & \textbf{0.060} & \textbf{0.922} \\
\textsc{AstroCo-L}    & \textbf{0.044} & \textbf{0.956} \\
\bottomrule
\end{tabular}
\end{table}

Table~\ref{tab:mainresults} reports reconstruction performance on MACHO R-band. 
Astromer v2 improves upon v1 with an RMSE of 0.113 ($R^2=0.73$), but both \textsc{AstroCo} variants substantially surpass these baselines: 
\textsc{AstroCo-S} achieves 0.060 ($R^2=0.922$) and \textsc{AstroCo-L} 0.044 ($R^2=0.956$).

\subsection{Downstream classification}
\begin{figure}[H]
    \centering
    \includegraphics[width=0.9\linewidth]{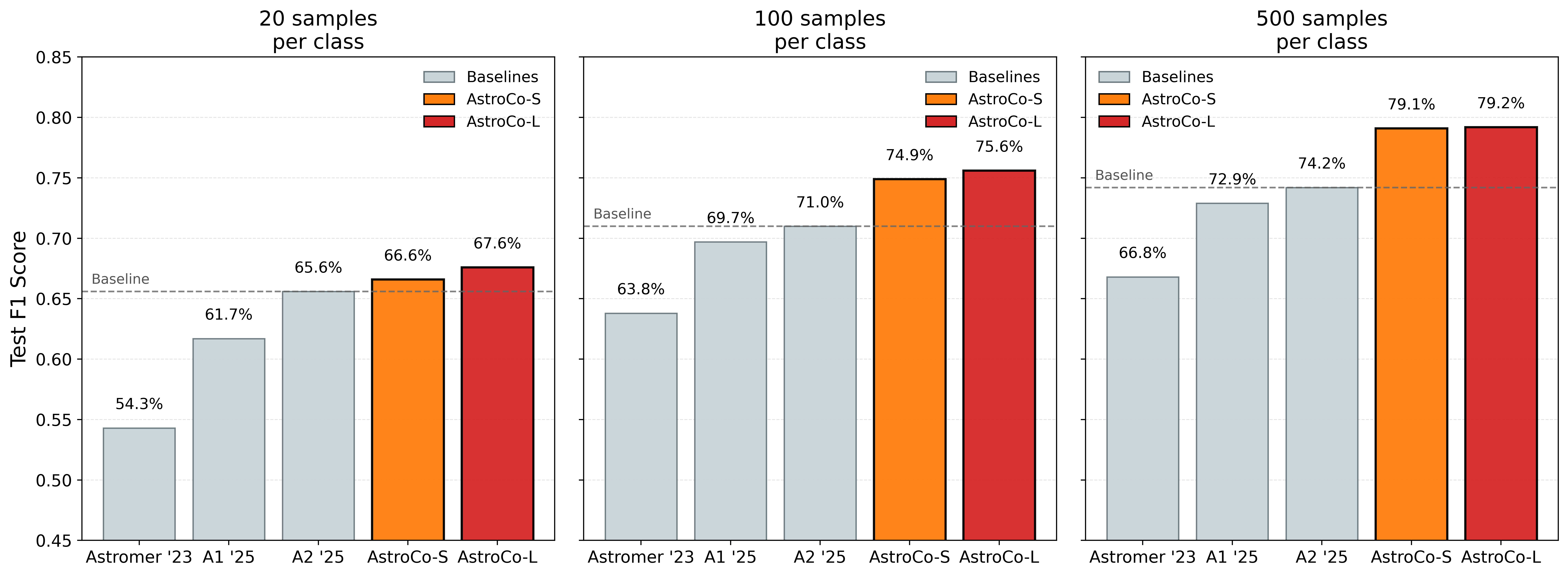}
    \caption{Downstream classification on the Macho-Labeled dataset \cite{Alcock2003}: macro-$F_1$ for 20/100/500 labels per class, averaged over 3 folds and 3 seeds. AstroCo-S/L (frozen encoder + linear head) outperform Astromer v1/v2, consistent with the results in Section~\ref{sec:results}.}
    \label{fig:alcock}
\end{figure}

Our frozen encoder + linear head outperforms Astromer v1/v2 on Alcock (20/100/500 shots)\cite{Alcock2003}; see Fig.~\ref{fig:alcock} for exact values.

\section{Conclusion and Future Work}\label{sec:conclusion}
\model{} integrates global self-attention for long-range dependencies with local depthwise convolution to capture short-term temporal patterns, enhanced by GLU-based gating and layer-wise mixing. Pretrained with masked reconstruction, it produces compact embeddings that excel in few-shot classification (20/100/500 labels per class), outperforming Astromer models in both reconstruction and classification under comparable budgets.

Future work includes extending to multi-band fusion, scaling across surveys, and exploring tasks such as anomaly detection and period estimation. We will also release pretrained weights and embeddings, whose quality and versatility make them valuable for label-efficient transfer, benchmarking, and broad community use across diverse astronomical applications.

\appendix

\section{Implementation Details}\label{app:implementation}
We implement \model{} in PyTorch, where each block combines MHSA, a depthwise Conv1D with $K{=}32$ (chosen after a sweep from $K=5$ to $K=128$ with no meaningful gains beyond $\sim K=32$), and a GLU-FFN with residuals and normalization.  
AstroCo-L uses $L{=}12$ layers, 4 heads of dimension 64 ($D{=}256$), $\sim$15.2M parameters, and AdamW with learning rate $7.5\times10^{-5}$.  
AstroCo-S uses $L{=}4$ layers, 4 heads of dimension 69 ($D{=}276$), $\sim$5.9M parameters, and learning rate $7.5\times10^{-4}$.  
Both variants use dropout 0.1, bf16, expansion ratio $r=4$, window size 200, and batch size 256.  
We apply early stopping with patience of 20 epochs. And we used zero scalar as masking token.

\section{Additional Experimental Details}
\subsection{Learning curves}\label{app:learning}
\begin{figure}[H]
    \centering
    \includegraphics[width=\linewidth]{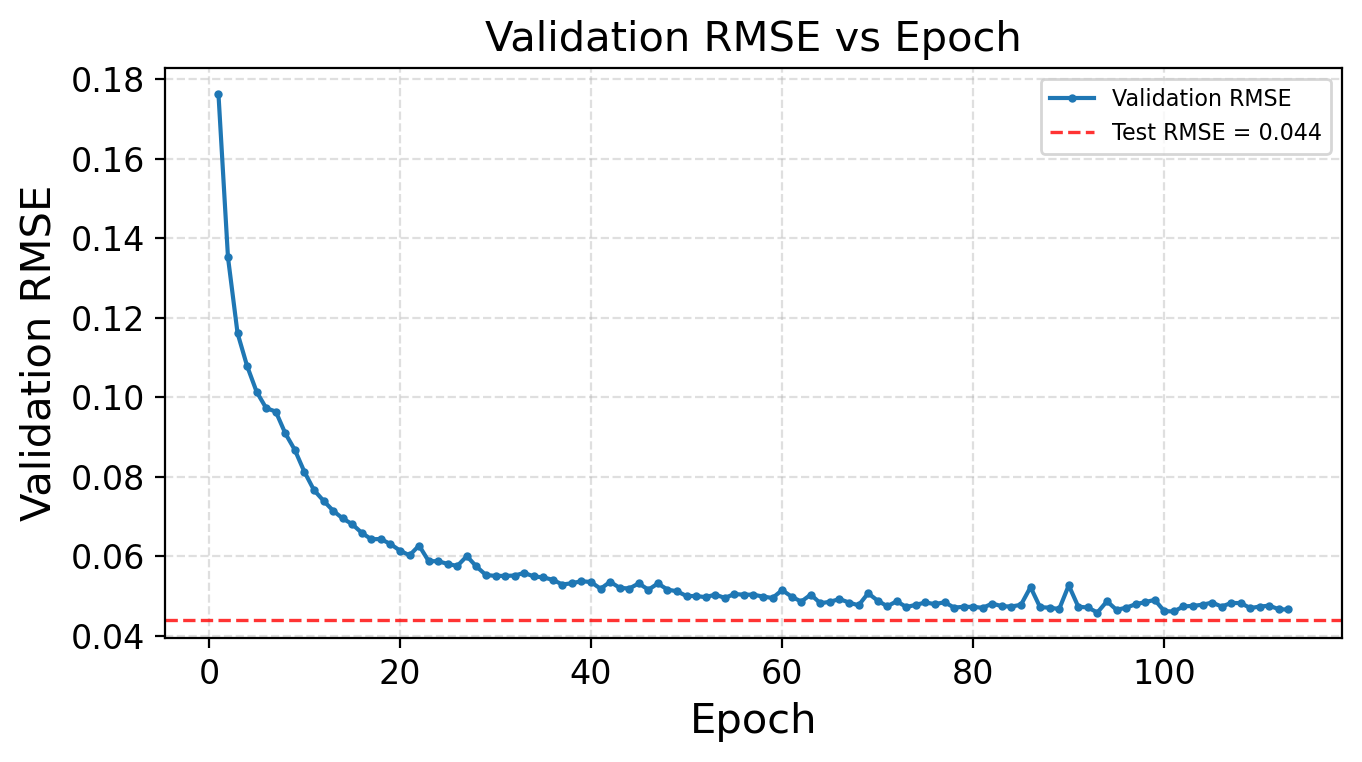}
    \caption{Validation reconstruction error over training epochs for \model{}-L, with reference to test RMSE.}
    \label{fig:learning_curve}
\end{figure}

Figure~\ref{fig:learning_curve} shows the learning dynamics of \model{}-L: validation RMSE steadily decreases across epochs and closely follows the test RMSE, indicating stable convergence and good generalization.

\subsection{Macho-labeled data set distribution}
\begin{table}[htbp]
\centering
\caption{Class distribution in the Macho-Label dataset \cite{Alcock2003}.}
\begin{tabular}{l l r}
\toprule
\textbf{Tag} & \textbf{Class Name} & \textbf{\# of Sources} \\
\midrule
Cep\_0 & Cepheid type I       & 1182 \\
Cep\_1 & Cepheid type II      & 683 \\
EC     & Eclipsing binary     & 6824 \\
LPV    & Long period variable & 3046 \\
RRab   & RR Lyrae type ab     & 7397 \\
RRc    & RR Lyrae type c      & 1762 \\
\midrule
Total  &                      & 20,894 \\
\bottomrule
\end{tabular}
\label{tab:alcock_classes}
\end{table}
\subsection{Training Objectives}\label{app:loss}
Given magnitudes $\{m_i\}_{i=1}^T$ and probed indices $\mathcal{M}\subseteq\{1,\dots,T\}$, 
the reconstruction loss is
\begin{equation}
  \mathcal{L}_{\mathrm{RMSE}}
  \;=\;
  \sqrt{\frac{1}{|\mathcal{M}|}\sum_{i\in\mathcal{M}}\big(\hat m_i - m_i\big)^2 }\,,
\end{equation}
where $\hat m_i$ are predictions from the regression head.  
For classification, we freeze the encoder and train a linear head with cross-entropy loss.

\end{document}